\documentclass[11pt]{article}
\usepackage{graphicx,amsmath,amssymb}
\newcommand{\ssy}[5]{#1  {\it #2}  {\bf #3} (#4) #5\rlap{.}}
\newcommand{\karti}[5][]{\begin{figure}[#2]\begin{center}%
\includegraphics[height=#3]{#4.ps}
\end{center}#1\caption{#5}\end{figure}}
\newcounter{rem}
\newcommand{\rem}{\refstepcounter{rem}\par\textbf{Remark.} }
\newcommand*{\ogr}[2]{#1\, \rule[-0.76em]{0.4pt}{1.4em} \,
\raisebox{-.55em}{$\displaystyle{}_{#2}$}  }
\newcommand{\rmd}{\mathrm{d}}
\DeclareMathOperator{\supp}{supp}
\title{Counter example to a quantum inequality}
\date{}
\author{S. Krasnikov}
\begin{document}
\maketitle
\begin{abstract}
A `quantum inequality' (a conjectured  relation between the energy density of
a free quantum field and the time during which this density is observed) has
recently been used to rule out some of the macroscopic wormholes and warp
drives. I discuss the possibility of generalizing that result to other similar
spacetimes and first show that the problem amounts to verification of a
slightly different inequality. This new inequality
\emph{can} replace the original one, if an additional assumption (concerning
homogeneity of the `exotic matter' distribution) is made, and \emph{must}
replace it if the assumption is relaxed. Then by an explicit example I show
that the `new'  inequality breaks down even in a simplest case (a free field
in a simply connected two dimensional space). Which suggests that there is no
grounds today to consider such spacetimes  `unphysical'.

\end{abstract}
\section{Introduction}
Suppose a freely falling observer measures the renormalized expectation value
of the energy density $\rho$ of a free quantum field in a state $\omega$.
Denote
\[
\overline{\rho}(\omega)
=\mathcal T^{-1}\int_{\tau_1}^{\tau_2}\rho(\omega;\tau)\,\rmd\tau,
\qquad\mathcal T\equiv|\tau_2-\tau_1|
\]
where the integral is taken along the observer's world line $\gamma$
parametrized by the proper time $\tau$. Obviously $\overline{\rho}$ is the
average energy density. The quantity we are interested in is
$\overline{\rho}\mathcal T^d$, where $d$ is the dimensionality of spacetime.
More specifically, we shall analyze the inequality
\begin{subequations}\label{eq:X}
\begin{equation}\label{eq:samoX}
-\overline{\rho}\mathcal T^d \lesssim 1
\end{equation}
(it is not trivial because we are concerned, in particular, with negative
$\rho$) under the condition
\begin{equation}\label{eq:T}
\mathcal T\lesssim\mathcal T_R\equiv
\big(\max |R_{\hat \imath\hat \jmath\hat m\hat n}|\big)^{-1/2}.
\end{equation}
\end{subequations}
In this expression the maximum  is taken over all $\tau\in (\tau_1,\tau_2)$
and all sets of indices. The hats over the indices mean that the components of
the Riemann tensor are found in the (orthonormal) frame associated with the
observer.

In the next section I discuss the importance of inequality~\eqref{eq:X} in
determining whether a spacetime is physically realizable. In
section~\ref{sec:couex} the massless scalar field is considered in the
two-dimensional de Sitter space. By explicit calculations the existence is
demonstrated of a `reasonable' state $\omega$ such that \eqref{eq:X} is
violated (in this letter I regard a state $\omega$ as reasonable if
$\langle\omega| T_{ab} |\omega\rangle^\text{ren}$ is smooth over the whole
spacetime).

\section{The quantum inequalities}\label{sec:disc}
Within the last decade dozens of papers (see \cite{RomMG} or \cite{thes} for a
review) have been published on inequalities that look similar to \eqref{eq:X}
and express the idea that ``the more negative the energy density is in an
interval, the shorter must be the duration of the interval"
\cite{RomMG,QICURV}. Unfortunately, all of them are often called equally ---
quantum inequalities (QI), which leads sometimes to confusion and even
mistakes. So, let us first establish terminology.

Pick a non-negative function $f$ normalized by
$$
 \int_{-\infty}^{-\infty} f(\tau)\,\rmd\tau=1
$$
and consider the `weighted average' of the  energy density
\[
\rho_f(\tau_0;\omega)\equiv \int_{-\infty}^{-\infty}
\rho(\tau;\omega)f(\tau-\tau_0)\,\rmd\tau
\]
(both integrals are taken along $\gamma$, see the Introduction). Denote by
$\mathcal F_{\omega_0}$ the Fock space built on a quantum state $\omega_0$.
Then \emph{difference} QIs  are the statements of the form
\begin{align*}\label{eq:I}
\rho_f(\omega) - \rho_f(\omega_0)\geq Q&
\qquad \forall\, \omega\in\mathcal F_{\omega_0},\tag{I}\\ %
\intertext{where $Q$  does not depend on $\omega$ and $\tau_0$. Their more
general variety (note the last quantifier) is}
\label{eq:II}
\rho_f(\omega) - \rho_f(\omega_0)\geq Q&
\qquad \forall\, \omega\in\mathcal F_{\omega_0},\
\forall\, \omega_0.\tag{II}\\ %
\intertext{A different type of QIs are those of the form}
\label{eq:III}
\rho_f(\omega)\geq Q&
\qquad\forall\,\text{reasonable } \omega.\tag{III}%
\end{align*}
Following \cite{Fewster} I shall call them \emph{absolute}.

Almost all quantum inequalities established so far are of type I. Among them
are, for example, those proved in
\cite{QUIN0} or in
\cite{difQI} (in the former $\omega_0$ is the standard Minkowski
vacuum $\omega_M$ and in the latter it is the vacuum defined by the timelike
Killing field present in the spacetime). The second group is represented by
the relation (4.1) of \cite{f}. The absolute QI have been derived so far only
for massless fields in two dimensions \cite{2dQI}--\cite{2dQIcF}, as is
 stated in~\cite{Fewster}.

The distinction between the difference and the absolute QIs is
\emph{fundamental}. In particular, \eqref{eq:I} does not become \eqref{eq:III}
even when $\rho_f(\omega_0)=0$, which happens in the Minkowski space when
$\omega_0=\omega_M$, and perhaps may also occur in curved spacetimes for some
special choice of $\omega_0$. The reason is that the corresponding inequality
holds not for the set of \emph{all} (reasonable) $\omega$, but only for a
\emph{very special subset} thereof (recall that $\omega$ can always be
transformed into $\omega'\notin\mathcal F_{\omega_0}$ by an appropriate
Bogolubov transformation). Thus a phrase like ``\dots arbitrarily extended
distributions of arbitrarily negative energy are not possible in Minkowski
spacetime''
\cite{QICURV} to be a \emph{correct} inference from such a QI
(the inequality \cite[(1)]{QICURV} in this case) has to be complemented by
quite a  restrictive stipulation: ``if the field is in the state
$\omega\in\mathcal F_{\omega_M}$.''
\rem The procedure proposed in \cite{Fewster} (the transition to the limit
$\Delta\to 0$, where $\Delta$ is the length of $\supp f$) transforms the
inequality  from \eqref{eq:I} into one looking \emph{similar} to that from
\eqref{eq:III} but  not satisfying an essential condition $Q\neq Q(\omega)$.
While the \emph{real} \eqref{eq:III} does not follow from  \eqref{eq:I} in
this limit either.

The importance of the QIs lies basically in the fact that they supposedly
``place some severe constraints on wormhole and warp drive spacetime
geometries" \cite{RomMG}. Specifically, it was argued
 that the `total negative energy' needed to maintain a human-size
Alcubierre bubble or Krasnikov tube is about
$10^{67}\,$g~\cite{Pfen97b,E&Ro97} and that the `absurdly benign' wormhole can
exist only if the exotic matter supporting it is concentrated in a
$10^{-68}\,$cm thick layer~\cite{portal}. These estimates are obtained in
assumption that for an appropriately chosen $f$ with $\supp f
\in (\tau_1,\tau_2)$
\begin{equation}\label{eq:qinOR}
\rho_f(\omega)> -{\mathcal T}^{-d}\quad\qquad\forall\,
{\mathcal T}\lesssim\mathcal T_R,\quad
\forall\,\text{reasonable } \omega.
\end{equation}
in any spacetime for any (free) quantum field\footnote{In the original papers
the assumption looked slightly different (in particular, $f$ had infinite
`tails'). This form it took a little later \cite{singed}. There is also an
additional requirement on ${\mathcal T}$, but\label{snoska} in simply
connected spacetimes without `boundaries' it is always satisfied.}. It should
be emphasized that
\eqref{eq:qinOR} is
\emph{a conjecture}. Being an absolute QI it, contrary to a common
misconception,
\emph{neither follows from, nor is supported by any of multiple difference
QIs}, as discussed above. At present it can be substantiated \emph{only} by
Flanagan and Vollick's QIs for massless fields in two dimensions
\cite{2dQI}--\cite{2dQIcF}.

Once the assumption \eqref{eq:qinOR} is made the reasoning goes, roughly, as
follows. Consider a point $p$ through which there is a timelike geodesic
segment $\gamma(\tau)$ of the length $\mathcal T_R$  with $\rho$ negative and
approximately constant along $\gamma$. Suppose that
\begin{equation}\label{eq:noC}
\max |R_{\hat \imath\hat \jmath\hat m\hat n}(p)|\approx
 \max |T_{\hat k\hat l}(p)|\approx -\rho(p)
\end{equation}
($p$ satisfying all these requirements happens to exist in all three
above-mentioned spacetimes). Applying
\eqref{eq:qinOR} with ${\mathcal T}=\mathcal T_R$ to $\gamma$ one can
write
\begin{equation}\label{eq:cep}
|\rho_f| <{\mathcal T}_R^{-4} = \big(\max |R_{\hat \imath\hat \jmath\hat m\hat
n}|\big)^{2}
\approx  \rho^2(p)
\end{equation}
or
\begin{equation}\label{eq:Pl_den}
|\rho(p)|\gtrsim \rho_f/\rho(p)\approx 1.
\end{equation}
Thus the energy density in $p$ must be Planckian! All the prohibitive values
for the `total amount of negative energy' that are interpreted as severe
restrictions on the warp drives stem just from this fact.

Those restrictions \emph{by themselves} are, however, of little interest: both
warp drives were designed to illustrate the idea of `faster-than-light' travel
and to do that by as simple means as possible. It seems not improbable that it
is just their simplicity  that leads to the undesirable properties of the
matter sources. That is why it would be important to find out to what
spacetimes the foregoing reasoning  can be generalized. A number of cases in
which it does
\emph{not} work are already known \cite{OluGr,portal}. Among them are
\begin{enumerate}\renewcommand{\theenumi}{\roman{enumi}}
  \item The spacetimes in which the `exotic matter' is generated by \emph{interacting}
quantum fields, because \eqref{eq:qinOR} \emph{is known} to break down in this
case, which is demonstrated with the Casimir effect;
  \item Some wormholes, because the condition
mentioned in the footnote~\ref{snoska} does not hold there;
  \item The
spacetimes in the relevant regions of which $\max |R_{\hat\imath\hat\jmath\hat
m\hat n}(p)|\gg
\max |T_{\hat k\hat l}(p)|$ for all observers (as, for example, in the vicinity of
\emph{any} gravitating body).
\end{enumerate}
And yet, a lot of interesting spacetimes are conceivable that should be
dismissed as `unphysical' be (an analog of) the restriction
\eqref{eq:qinOR}  true.

Thus the quantum inequality \eqref{eq:qinOR} deserves to be tested most
seriously. In doing so, however, one has to specify $f$ more definitely,
because generally $\rho_{f_1}(\omega)\neq \rho_{f_2}(\omega)$ (unless, of
course, exotic matter is distributed homogeneously as in the example above,
but no reasons are seen to restrict ourselves to this case). And it is quite
possible that among QIs obtained from \eqref{eq:qinOR} by specifying $f$  some
are true and some other are not. For example, one can choose $f$  to be the
characteristic function of the interval $(\tau_1,\tau_2)$:
\begin{equation*}\label{eq:char}
f=h\colon\qquad  h(\tau)= 1\quad
\text{for }\tau\in (\tau_1,\tau_2),
\qquad h(\tau)=0\quad  \text{otherwise}.
\end{equation*}
$\rho_f$ then is just $\rho_h=\overline\rho$ and the inequality in
\eqref{eq:qinOR} transforms into \eqref{eq:X}.
As well one could choose another $f$:
\begin{equation}\label{eq:FV} f\in C^\infty, \qquad
  \int_\gamma{f'}^2f^{-1}\,\rmd\tau<1.
\end{equation}
and get another inequality\footnote{The similarity of these two inequalities
does not, of course, make them the \emph{same}. It is misunderstanding of this
fact that underlies the recent paper \cite{Fewster}. For example in the key
statement of the paper ``we have shown that Krasnikov's supposed
counterexample to the QEIs is, in fact, entirely compatible with them"  the
word `QEIs' refers (or, at least, \emph{must} refer) to  inequality
\eqref{eq:X}, while the word `them' --- to  the inequality from
\eqref{eq:qinOR} with $f$ satisfying \eqref{eq:FV}.}.
The state considered in the next section satisfies \eqref{eq:qinOR},
\eqref{eq:FV} (in fully agreement with the results of
\cite{2dQIcF}) and at the same time, as we shall see,
violates \eqref{eq:X}.

Fortunately, one does not need to test \emph{all} possible $f$. The point is
that the whole reasoning cited above fails when so does \eqref{eq:X}. To see
this consider a situation in which  \eqref{eq:X} breaks down, while
\eqref{eq:qinOR} with $f=g\neq h$ does not. That clearly would
imply
\begin{equation}\label{eq:wevsunw}
|\rho_g|\ll |\overline\rho|.
\end{equation}
The approximate equality in  \eqref{eq:cep} to hold it is necessary that
$|\rho(p)|\approx \max_\gamma|\rho|$ and, as a consequence,
$|\overline\rho|\lesssim |\rho(p)|$. The latter inequality combined with
\eqref{eq:wevsunw} implies $|\rho_g|\ll|\rho(p)|$. So,
instead of \eqref{eq:Pl_den} one gets
\[
|\rho(p)|\gtrsim \rho_g/\rho(p)\ll 1.
\]
Quite a useless result!

To summarize, if the inequality \eqref{eq:X} held for all reasonable quantum
states it would  mean that a wide class of `shortcuts' \cite{portal} require
Planck scale energy densities. \eqref{eq:X} in this implication can be
substituted \emph{neither} by a difference QI \emph{nor} by the system
\eqref{eq:qinOR},\eqref{eq:FV}\footnote{Though if one day the latter is
proven it will rule out the spacetimes in which exotic matter of moderate
density is distributed homogeneously.}.

\section{The counter example}\label{sec:couex}

That \eqref{eq:samoX} can be violated even in the Minkowski space follows
immediately from the results of \cite{2dQI}. However, the corresponding state
would have singularities at $\tau_{1,2}$. To construct a `reasonable' state
violating \eqref{eq:X} consider the two-dimensional de Sitter spacetime. One
of the customary choices of coordinates casts the metric into the form
\begin{equation*}
\rmd s^2=\frac{\alpha^2}{\sin^2(\mu-\nu)}\,\rmd\mu\rmd\nu,
\qquad (\mu+\nu)\in \text{I\!R},\ (\mu-\nu)\in(0,\pi).
\end{equation*}
However, to fix in a convenient way the quantum state of the field we
shall use different coordinates. Pick a function $w$ such that
\begin{equation*}
  w\in C^\infty(\text{I\!R}),\quad
w'\neq 0,\quad
\ogr{w(x)}{(\epsilon,\frac{\pi}{2}-\epsilon)}=\tfrac{1}{2}\ln\tan x,
\end{equation*}
where $\epsilon$ is a small positive constant. [For example, $w$ can
be defined as
\[
w(x)=\int_{\frac{\pi}{4}}^x [\chi_1(s)+ \chi_2(s)\sin^{-1}2 s]\,\rmd s,
\]
where $\{\chi_i\}$ is a partition of unity for the cover $\{X_i\}$ of
I\!R, with $X_1=(-\infty,\epsilon)\cup(\frac{\pi}{2}-\epsilon,\infty)$, and
$X_2=(\frac{\epsilon}{2},\frac{\pi-\epsilon}{2})$.]
The new coordinates are defined as follows
\[
u(\mu)=w(\mu),\quad v(\nu)=w(\nu),\qquad
t= \tfrac{1}{2}(v-u),\quad x= \tfrac{1}{2}(v+u)
\]
and the metric takes the form
\begin{subequations}\label{eq:C}
\begin{equation}
\rmd s^2= C\rmd u\rmd v,
\end{equation}
where $C$ is smooth and nonzero in the whole space. In the region $W$, see
figure~\ref{fig:ris2},
\karti{tb}{0.3\textwidth}{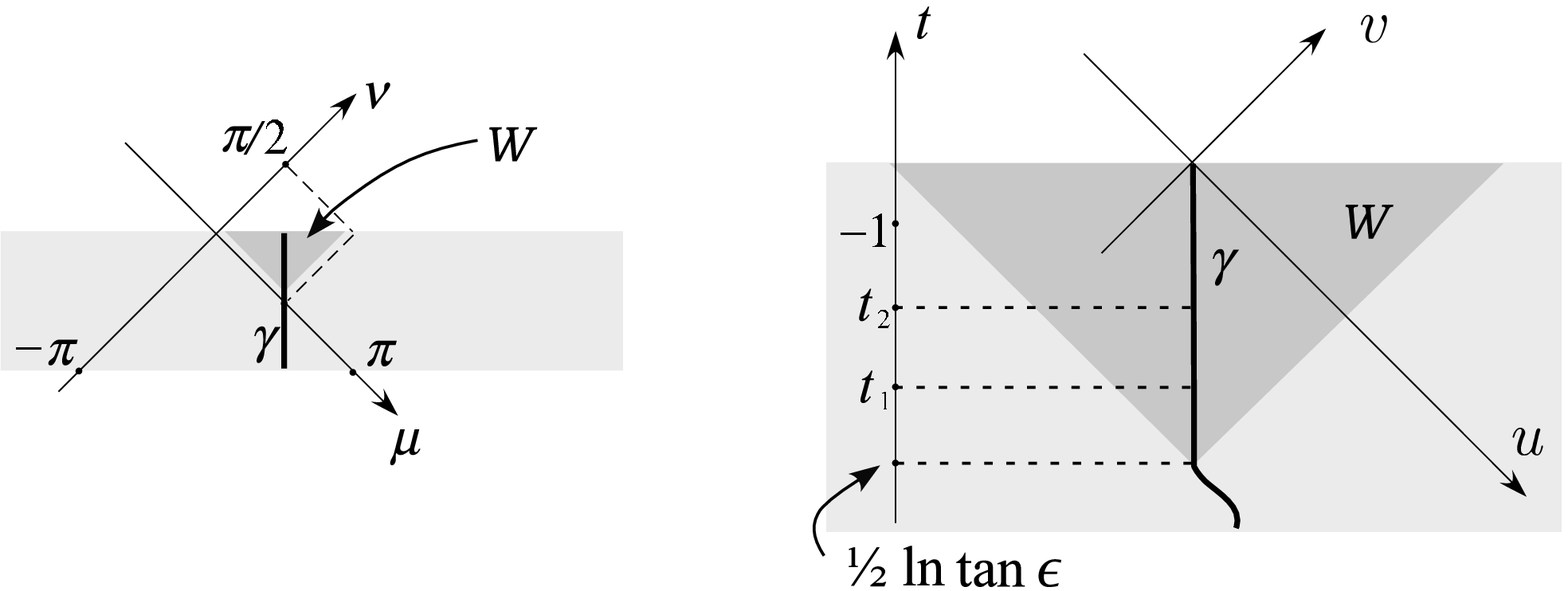}{\label{fig:ris2} The region $W$ (dark
gray) of the de Sitter space (light gray) in different coordinates.}
\[
W\colon\qquad
\epsilon<\mu,\nu<\tfrac{\pi}{2}-\epsilon,
\quad \nu<\mu,
\]
$C$ can be expressed explicitly:
\begin{multline}
C=\frac{\alpha^2}{\sin^2(\mu-\nu)}
\frac{\rmd \mu}{\rmd u}\frac{\rmd \nu}{\rmd v}\ogr{}{W}
\\
=\frac{4\alpha^2\cos^2\nu\cos^2\mu}{
\sin^2(\mu-\nu)\cdot e^{-2u}e^{-2v}}=
\alpha^2\sinh^{-2}(u-v).
\end{multline}
\end{subequations}

Consider now the massless  scalar field in the conformal vacuum state
associated with the metric \eqref{eq:C}. Its renormalized stress-energy tensor
in $W$ is readily found by formula (6.136) of \cite{BirDav}:
\begin{equation*}
\begin{aligned}
  T_{vv}= T_{uu} & =-\tfrac{1}{12\pi}C^{1/2}\partial_u^2C^{-1/2}=
-\tfrac{1}{12\pi}.
\\
  T_{uv} &=\tfrac{1}{96\pi}RC=
   \tfrac{1}{12\pi}\sinh^{-2}(u-v)
  \end{aligned}
\end{equation*}
Let $\gamma\subset W$ be a geodesic segment
\[
x=0,\quad t\in[t_1,t_2],
\]
where $t_{1,2}$ are chosen so that
\begin{equation}\label{eq:por-k_t}
\tfrac{1}{2}\ln\tan\epsilon<t_1<t_2\ll -1.
\end{equation}
For an observer whose world line is $\gamma$ the energy density is\footnote{I
am grateful to Dr.~Fewster for the detailed evaluation \cite{Fewster} of the
coefficient (it is 1 and not $-1/2$ as was erroneously written in the draft of
this letter) with which the term $\sinh^{-2}2t$ enters the expression. This
coefficient, though, plays absolutely no role in what follows as we  neglect
the whole term [on the strength of the condition \eqref{eq:por-k_t}].}
\[
\rho=T_{\hat 0\hat 0}=
T_{tt}C^{-1}
=2(T_{uu}-T_{uv})C^{-1}=-\tfrac{1}{6\pi}(1+\sinh^{-2}2t)C^{-1}
\approx -(6\pi C)^{-1},
\]
which
yields
\[
-\overline\rho=\frac{\mathcal T^{-1}}{6\pi}\int^{t_2}_{t_1}
C^{-1}
C^{1/2}\,\rmd t \approx
\frac{\mathcal T^{-1}}{24\pi}\alpha^{-1} e^{-2t_1}.
\]
At the same time
\[
\mathcal T=-\int^{t_2}_{t_1}\,\alpha\sinh^{-1}2t\,\rmd t
\approx \alpha\int^{2t_2}_{2t_1}e^{s}\rmd s
\approx
\alpha e^{2t_2}.
\]
Thus,
\[
-\overline\rho{\mathcal T}^2=\tfrac{1}{24\pi}e^{2(t_2-t_1)},
\]
which obviously --- just by picking sufficiently small $\epsilon$
and $t_1$ --- can be made arbitrarily large in contradiction with
\eqref{eq:X} [note that
\[
\mathcal T\ll \mathcal T_R=\alpha
\]
and so the criterion \eqref{eq:T} is satisfied].

\section{Conclusions}
In a number of papers warp drives and wormholes have been labelled as
unphysical because of the enormous energy densities supposedly needed for
their existence\footnote{Ford and Roman argue that a large discrepancy in the
length scales characterizing these spacetimes (if it takes place)  would also
be grounds for rejecting them as unphysical. But in the absence of explanation
this thesis does not look convincing. The stars, for example,  are nothing but
associations of protons (and electrons), which are characterized by the length
scale $r_\text{pr}\sim 10^{-15}\,$m. So, Deneb, say, with its radius
$R_\text{D}\gtrsim 10^{11}\,$m is characterized by the length scales with the
ratio $R_\text{D}/r_\text{pr}\gtrsim 10^{26}$. The ratio of the mass scales is
$M_\text{D}/m_\text{pr}\gtrsim 10^{58}$. Is Deneb unphysical?}. This idea is
supported by the results obtained for the Alcubierre bubble and Krasnikov tube
and based on the conjecture
\eqref{eq:qinOR},\eqref{eq:FV}. The generalization of these results, however,
would require the validity of a slightly different inequality, namely
\eqref{eq:X}, for all reasonable quantum states in
reasonable spacetimes. In fact, however, the inequality turns out to be
invalid even in a simplest case of the massless scalar field in the two
dimensional de Sitter space. Consequently, there are no reasons any longer to
expect that the shortcuts are necessarily connected with Planckian energy
densities. So, these objects are perhaps not that unphysical after all.

\section*{Acknowledgements}
I wish to thank the members of the Friemann seminar and particularly
Yu.~Pavlov for helpful discussion.

\end{document}